\documentclass[Physsubmission, Phys]{SciPost}

\hypersetup{
    colorlinks,
    linkcolor={red!50!black},
    citecolor={blue!50!black},
    urlcolor={blue!80!black}
}

\usepackage[bitstream-charter]{mathdesign}
\DeclareSymbolFont{usualmathcal}{OMS}{cmsy}{m}{n}
\DeclareSymbolFontAlphabet{\mathcal}{usualmathcal}

\newcommand{\tvec}[1]{\vec{#1}_\bot}

\renewcommand\r{\rho}

\newcommand\e{\epsilon}

\newcommand\m{\mu}

\newcommand\x{\xi}

\newcommand\s{\sigma}

\newcommand\G{\Gamma}

\renewcommand{\vec}{\boldsymbol}
\renewcommand{\part}{{\rm part}}

\newcommand{\be}{\begin{equation}}
\newcommand{\ee}{\end{equation}}
\newcommand{\bes}{\begin{subequations}}
\newcommand{\ees}{\end{subequations}}
\newcommand{\bea}{\begin{eqnarray}}
\newcommand{\eea}{\end{eqnarray}}

\newcommand{\pa}{\partial}

\begin{document}

\begin{center}{\Large \textbf{
Jets in Evolving Matter within the Opacity Expansion Approach
\\
}}\end{center}

\begin{center}
Andrey V. Sadofyev\textsuperscript{1, \,2$\star$},
Matthew D.  Sievert\textsuperscript{3} and
Ivan Vitev\textsuperscript{4}
\end{center}

\begin{center}
{\bf 1} Instituto Galego de F{\'{i}}sica de Altas Enerx{\'{i}}as,  Universidade de Santiago de Compostela, Santiago de Compostela 15782,  Spain
\\
{\bf 2} Institute of Theoretical and Experimental Physics, NRC Kurchatov Institute, Moscow 117218, Russia
\\
{\bf 3} Department of Physics, New Mexico State University, Las Cruces, NM 88003, USA
\\
{\bf 4} Theoretical Division, Los Alamos National Laboratory, Los Alamos, NM 87545, USA
\\
*andrey.sadofyev@usc.es
\end{center}

\begin{center}
\today
\end{center}


\section*{Abstract}
{\bf
In a recent study \cite{Sadofyev:2021ohn} we have extended the opacity expansion approach to describe jet-medium interactions including  medium motion effects in the context of heavy-ion collisions.  We have computed color field of the in-medium sources, including the effects of the transverse field components and the energy transfer between the medium and jet. The corresponding contributions are sub-eikonal in nature, and were previously ignored in the literature.  Here we discuss how our approach can be applied to describe the medium motion effects in the context of Deep Inelastic Scattering.
}

\section{Introduction}
\label{sec:intro}
All  first-principles approaches that describe jet-medium interactions,  both for cold and hot nuclear matter,  start by characterizing the medium using a collective color field, see \cite{Sadofyev:2021ohn} and references therein.  This color field can be considered to be sourced by (quasi)particles of the nuclear matter, including the strong classical fields associated with gluon saturation in the color-glass condensate (CGC) effective theory.  The interaction process, in turn,  is commonly described in the so-called eikonal limit -- the limit of infinite energy of the leading parton.  Then the problem is considerably simplified,  since the transverse field components are suppressed by the jet energy comparing to the longitudinal and temporal ones. An additional simplification comes from a kinematic constraint on the momentum transfer.  For instance, in the case of infinitely massive static sources,  the energy transfer is zero. 

These assumptions  make the calculations tractable,  and are natural for highly energetic jets in heavy-ion collisions (HIC). However, in this limit jets are completely decoupled from the transverse medium evolution. Thus,  if one  attempts to use jets for a tomographic study of nuclear matter, the assumptions above should be unavoidably revisited.  In \cite{Sadofyev:2021ohn} it was shown that the leading flow effects can be accounted for already at the level of the first sub-eikonal correction.  Moreover, the formalism developed in \cite{Sadofyev:2021ohn} can be applied to describe the effects of the medium inhomogeneity in the transverse directions,  making one more step towards the goal of the jet tomography. 

Here we argue that the very same approach can be generally applied to describe the jet-medium interactions in other forms of nuclear matter,  including the non-equilibrium CGC state produced in HIC and cold nuclear matter at the future Electron-Ion Collider (EIC).

\section{The Opacity Expansion in an Evolving Matter}
\label{sec:Theory}

If the QCD medium is made of color (quasi)particles of mass $M$, then its collective color field can be constructed from the fields of separate sources $A_\mathrm{ext}^{\mu a} (q) = \sum_i e^{i q \cdot x_i} a_i^{\mu a} (q)$. Notice that the phase factors involve only the spatial positions of the sources at the initial time. Following \cite{Sadofyev:2021ohn} we assume that the matter and jet consist of scalar particles, while they interact through $t$-channel exchanges of standard QCD gluons.\footnote{This replacement simplifies the calculation while leaving the medium motion corrections untouched in the regime of interest, see \cite{Sadofyev:2021ohn}.} Then, the form of an elementary field produced by a single source $i$ reads
\begin{align}
\label{e:potl2}
    a_i^{\mu a} (q) = (i g \, t_i^a) \, (2 p_{s \, i} - q)_\nu
    \left( \frac{-i g^{\mu \nu}}{q^2 - \mu_i^2 + i \epsilon} \right)
    \: (2\pi) \, \delta\Big( (p_{s \, i} - q)^2 - M^2 \Big) \, ,
\end{align}
where $t_i^a$ is the $SU(N_c)$ generator in the appropriate representation in the color space of the medium particle $i$, $p_{s\,i}$ is its momentum, $g$ is the coupling, and $\m_i$ is the Debye mass of the t-channel gluon at the position of $i$th source. Taking the limit of heavy sources, we find
\begin{align}
\label{e:potl4}
    g \, a_i^{\mu a} (q) &= 
    t_i^a \, u_i^\mu \: v_i (q) \: 
    (2\pi) \, \delta\left( q^0 - \vec{u}_{i} \cdot \vec{q} \right) \, , 
\end{align}
where $u^\m_i=p_{s\,i}^\m/p^0_{s\,i}$ is the nonrelativistic velocity, and $v(q)\equiv\frac{-g^2}{\vec{q}^2+\m^2-q_0^2-i\e}$ is the Gyulassy-Wang potential.
%
\begin{figure}
    \centering
	\includegraphics[width=0.4\textwidth]{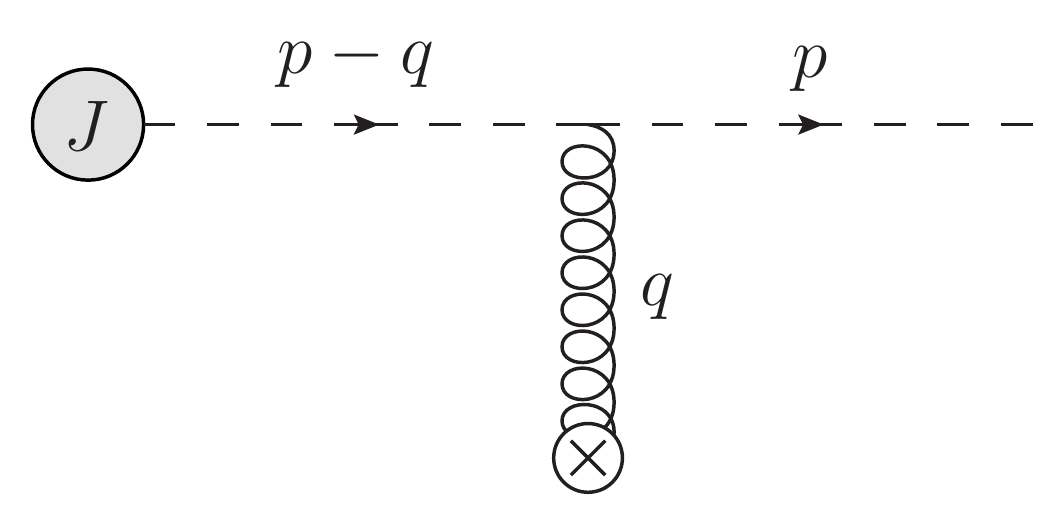}
	\includegraphics[width=0.4\textwidth]{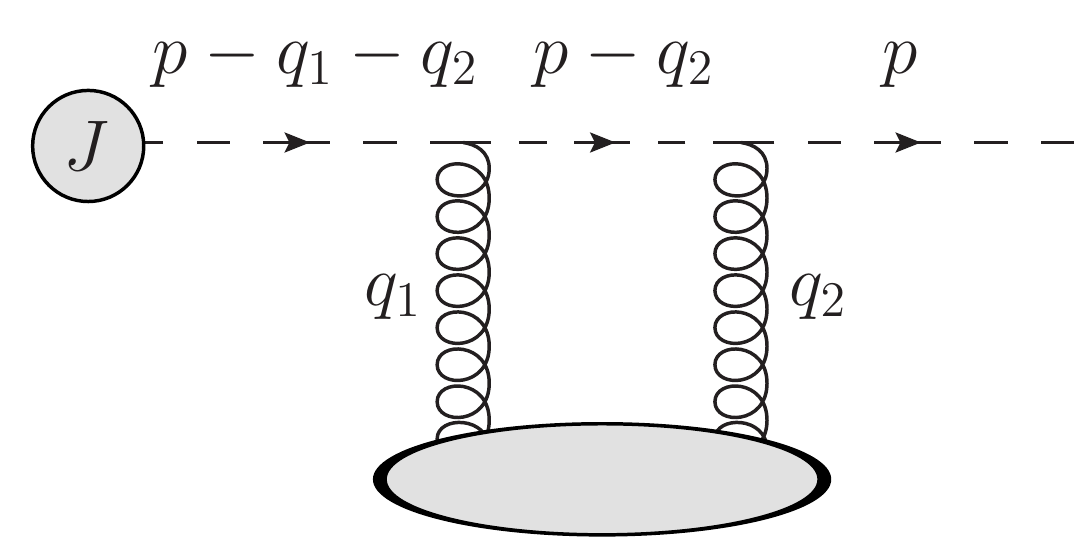}
	\caption{The single-Born (SB) amplitude $M_1$ (left) and double-Born (DB) amplitude $M_2$ (right) for transverse-momentum broadening in an external potential.}
	\label{f:Broad1}
\end{figure}

Now we are ready to consider the simplest jet-medium interaction process -- jet broadening. For this purpose, we introduce an initial distribution of jets $E \, \frac{dN^{(0)}}{d^3 p} \equiv \frac{1}{2(2\pi)^3} \left| J(p) \right|^2$ produced by a hard-scattering process at $x_0=0$, and such that $p_z\approx E$. At the first order in the opacity expansion, this process involves only the two diagrams shown in Fig. \ref{f:Broad1}. The single Born 
(SB) amplitude can be explicitly written as
\begin{align}
\label{e:BroadBorn1}
    i M_1 (p) = \int \frac{d^4 q}{(2\pi)^4}
    \Big[ i g \, t^a_\mathrm{proj} A_\mathrm{ext}^{\mu a} (q) \: (2 p - q)_\mu \Big]
    \left[  \frac{i}{(p-q)^2 + i \epsilon}  \right] J(p-q) \, ,
\end{align}
where $t^a_\mathrm{proj}$ is the $SU(N_c)$ generator of the jet parton and $p^\mu\approx\Big( E \, , \, \vec{p}_\bot \, , \, E - \tfrac{p_\bot^2}{2 E} \Big)$ in the eikonal expansion. The momentum $q$ enters (\ref{e:BroadBorn1}) through the interaction vertex, with non-zero energy transfer controlled by the delta function \eqref{e:potl4}. Later this energy transfer propagates into the jet source function $J(p-q)$, potential $v(q)$, and propagator.

The amplitude should be squared and averaged over the quantum numbers. Here, another essential assumption enters -- a color-neutrality condition, which can be represented as $\left\langle t^a_i t^b_j \right\rangle\equiv\frac{1}{2  C_{\bar{R}}} \delta_{i j} \delta^{a b}$, with $C_{\bar{R}}$ being the quadratic Casimir in the representation opposite to the representation of the in-medium source. Replacing the sum over the sources with an integral $\sum_i f_i =\int d^3 x \: \rho(\vec{x}) \: f(\vec{x})$, where $\r(\vec{x})$ is the source density, we write
\begin{align}
\label{e:BroadBorn10}
    \left\langle \left| M_1 \right|^2 \right\rangle &= 
    \mathcal{C}\,\int d^3 x \:
    \frac{d^2q_\bot} {(2\pi)^2}  \frac{d^2 q'_\bot}{(2\pi)^2} \, \rho(\vec{x}) \:
    \exp\left[ -i\:\frac{\vec{u}_{\bot} \cdot (\vec{q}_{\bot} - \vec{q}_{\bot}^{\: \prime})}{1-u_z}z- i \: \frac{(p-q)_\bot^2 - (p-q')_\bot^2}{2 E (1-u_z)} z\right] \:
    \notag \\ & \hspace{-0.5cm} \times
    e^{-i (\vec{q}_\bot - \vec{q}_\bot^{\: \prime}) \cdot \vec{x}_{\bot}} \:v (q_\bot^2) v^* (q_\bot^{\prime \, 2}) \: J(E, \vec{p}_\bot - \vec{q}_\bot) J^*(E, \vec{p}_\bot - \vec{q}_\bot^\prime)
    \Bigg[ 1 
    + \vec{u}_{\bot} \cdot \vec{\Gamma}_\bot (\vec{q}_\bot, \vec{q}_\bot^\prime)
    \Bigg] \, .
\end{align}
Here we have introduced a shorthand notation for the overall color factor $\mathcal{C} \equiv \frac{C_\mathrm{proj}}{2 C_{\bar{R}}}$, with $C_\mathrm{proj} \mathbf{1}= t^a_\mathrm{proj} t^a_\mathrm{proj}$.  The whole sub-eikonal correction is now separated into a single structure $\vec{\Gamma} (\vec{q}_\bot , \vec{q}_\bot^{\: \prime})$; the general form is beyond the scope of this discussion, but
if the medium parameters are $\vec{x}_\perp$-independent, then we can replace $\int d^2 x_\perp \, e^{-i (\vec{q}_\bot - \vec{q}_\bot^{\: \prime}) \cdot \vec{x}_{\bot}}$ with a delta function, giving
\begin{align}
    \label{e:GamNew}
     \vec{\Gamma} (\vec{q}_\bot)&=
    - 2 \frac{\vec{p}_\bot - \vec{q}_\bot}{(1-u_{z})E}
    + \frac{\vec{q}_\bot}{(1-u_{z}) E} \:
    \frac{(p-q)_\bot^2 - p_\bot^2}{\bar\sigma(q_\bot^2)} \: 
    \frac{\pa\bar\sigma}{\pa q_\bot^2}
    - \frac{\vec{q}_\bot}{1-u_z} \frac{1}{|J (E, \vec{p}_\bot - \vec{q}_\bot)|^2} \frac{\partial |J|^2}{\partial E} \,,
\end{align}
where $\bar{\sigma}(\vec{q}_\bot) \equiv \frac{d\sigma}{d^2q_\bot} = \frac{1}{(2\pi)^2} \mathcal{C} |v(q_\bot^2)|^2$.

The double Born (DB) amplitude can be similarly evaluated \cite{Sadofyev:2021ohn}, and the jet momentum distribution at the 1st order in opacity reads
\begin{align}
\label{e:untrt1}
E \frac{dN^{(1)}}{d^3 p} 
    &= \int dz \, d^2q_\bot \,\r(z)\: {\bar\s}(q^2_\perp)\Bigg[\left( E \frac{dN^{(0)}}{d^2 (p-q)_\perp \, dE} \right) \bigg( 1
    + \vec{u}_{\bot} (z) \cdot  \vec{\Gamma} (\vec{q}_\bot) \bigg)\notag\\
    & \hspace*{4cm} - \left( E \frac{dN^{(0)}}{d^2 p_\perp \, dE} \right) \bigg( 1 + \vec{u}_{\bot} (z) \cdot  \vec{\Gamma}_{DB} (\vec{q}_\bot) \bigg)\Bigg]\,,
\end{align}
where $\G_{DB}$ is the corresponding sub-eikonal contribution. Taking a simple model for the initial profile $E \frac{dN^{(0)}}{d^3 p} = \frac{1}{2(2\pi)^3} | J(p) |^2 \propto E^{-4} \, \delta^{(2)} (\tvec{p})$, we notice that the DB contributions decouple. We further assume the medium properties to be $z$-independent, and find
\begin{align}   \label{e:moments2}
    \langle \vec{p}_\perp (p_\perp^2)^k \rangle &=
    \frac{\tvec{u}}{(1-u_z)} \frac{L}{\lambda}  \frac{(\mu^2)^{k+1}}{E}
    \int_0^\infty \,d\xi\, \frac{\xi^{k+1}(2+3\x)}{(1+\xi)^2}\,.
\end{align}

If the matter is inhomogeneous in the transverse directions, the averaging procedure becomes pretty involved. Following the idea of the hydrodynamic gradient expansion\footnote{See \cite{Lekaveckas:2013lha, Rajagopal:2015roa, Sadofyev:2015hxa, Sadofyev:2015tmb, Reiten:2019fta} for recent applications of the gradient expansion in the context of probe-medium interactions.}, we assume the transverse gradients to be small. Then the $x_\perp$-integral can be again performed analytically, and for the leading gradient effects at zero velocity, the distribution reads
\begin{align}
\label{e:untrt2}
    &\left( E \frac{dN^{(1)}}{d^3 p} \right)^\mathrm{(linear)}= 
    \int dz\, \int d^2 q_\bot\,\bar\s(q_\bot^2)\left(\pa^j\r+\r\,\frac{1}{\bar\s(q_\bot^2)}\frac{\pa\bar\s}{\pa\m^2}\,\pa^j\m^2\right)\notag\\
    &\hspace{3.5cm} \times \left\{\left( E \frac{dN^{(0)}}{d^2 (p-q)_\perp\, dE} \right)\left[\frac{(p-q)_\perp^j}{E}z\right]-\left( E \frac{dN^{(0)}}{d^2p_\perp\, dE} \right)\left[\frac{p_\perp^j}{E}z\right]\right\} \, .
\end{align}
The moments of this distribution are zero unless the initial profile has a finite width. For a simple model profile $E \frac{dN^{(0)}}{d^3 p} = \frac{f(E)}{2\pi w^2} \, e^{-\frac{p_\perp^2}{2w^2}}$ with width $w$, one finds the linearized gradient effect
\begin{align}   
    \langle \vec{p}_\perp \, p_\perp^2 \rangle^{(linear)}&\simeq 
    \frac{L}{\lambda} \, \frac{L}{E} \,
    w^2 \mu^2
    \frac{\tvec{\nabla}\rho}{\rho} \, 
    \ln\frac{E}{\mu}\,,
\end{align}
indicating a non-trivial interplay between different contributions to the odd moments of the jet momentum distribution. A direct check shows that the even moments are unmodified by the flow velocity and gradients in the considered regimes.

\section{Outlook and Conclusions}
One powerful advantage of the formulation in Sec.~\ref{sec:Theory} of jet-medium interactions as energetic partons propagating in a background field is that this formalism can be applied equally well both to the hot nuclear matter produced in HIC and to cold nuclear matter as probed in deep inelastic scattering as in the future EIC.  The opacity expansion formalism as we use it here has been applied in this context to describe the propagation of jets through heavy nuclei at the EIC, see e.g., Ref.~\cite{Sievert:2018imd} and references therein.  Although the mechanism by which jets couple to the background field -- including the corrections due to medium gradients and motion -- is universal, the nature of that background field and the information it carries will change substantially between hot and cold nuclear matter.

At its most fundamental level, the first-order opacity expressions for jet modification reflect the full correlations of four partonic fields evaluated in the target state: two fields describing the hard scattering which produces the jet, and two gluonic fields describing the first-order-opacity rescattering of the jet in the medium.  Thus, without performing any medium averaging, the information encoded in the jet-medium interactions is described by a correlator such as $\langle A \, A \, A \, A \rangle$ (if the initial hard scattering were mediated by a gluon, in this example).  This general correlator is associated with a twist-4 parton distribution in the collinear limit; while an all-orders proof of twist-4 factorization is currently lacking, there has been important work exploring the connection between twist-4 operators and momentum broadening in the opacity expansion, see, e.g. Ref.~\cite{Kang:2014ela} and references therein.  

A single twist-4 distribution associated with the operator $\langle A \, A \, A \, A \rangle$ is appropriate for jet interactions with a proton target, where all degrees of freedom including color may be fully correlated.  If the target is a heavy nucleus, however, this introduces a larger length scale $L \propto A^{1/3}$ from the length $L$ of the nucleus (or equivalently its mass number $A$).  Then the dominant contributions are those which are length-enhanced; since color correlations are limited to a confinement scale $\sim 1/\Lambda_{QCD} \ll L$ this leads to a factorization of the generic matrix element into pairs of well-separated operators, $\langle A \, A \, A \, A \rangle \sim \langle A \, A \rangle \: \langle A \, A \rangle$, which reflect the spatial correlations among the densities of partons, but not their color degrees of freedom.  Relationships of this type between collinear parton distributions and final-state interactions as in jet physics were explored in, e.g., Refs.~\cite{Kang:2014ela} and \cite{Qiu:2004da}, such as the relationship between the jet-medium parameter $\hat{q}$ and twist-4 parton densities.  

In terms of the velocity corrections discussed here, if one substitutes the velocity field $u^\m_i=p_{s\,i}^\m/p^0_{s\,i}$ back from its definition, then these important corrections reflect information about the distribution of parton momenta inside the target system.  In a proton, this will correspond to a particular four-field matrix element describing the color currents of partons.  In a large system like a heavy nucleus where the color correlations decouple, the jet will interact with the distribution of momentum inside the target.  The distribution of orbital angular momentum within cold nuclear matter is a central question for the future EIC, and previous work has shown that the orbital motion of cold nuclear matter can be identified with the transverse-momentum-dependent parton distributions of nuclei \cite{Kovchegov:2015zha}.  This need not apply only to heavy nuclei; any QCD system at sufficiently high energies dynamically generates its own semi-hard color screening scale (the saturation momentum $Q_s$) described by the CGC effective theory, see e.g. Ref.~\cite{Kovchegov:2012mbw} and references therein.  This again leads to a factorization of the general correlator into pairwise densities as before; many of the same techniques have also been successfully applied to studies of gluon saturation in this framework, e.g. Ref.~\cite{Kovner:2017ssr}.  

Thus, by extending the formalism presented here to cold nuclear matter, one can use jets as probes to study correlations among multi-parton distributions, color currents, and orbital angular momentum in cold nuclear matter.

\paragraph{Funding information}
The work of A.V\hspace{0mm}.S. on Secs. I and II is supported by the Russian Science Foundation Grant RSF 21-12-00237 and on Sec. III by the European Research Council project ERC-2018-ADG-835105. A.V\hspace{0mm}.S. is also grateful for support from Xunta de Galicia (Centro singular de investigaci\'on de Galicia accreditation 2019-2022),  from the European Union ERDF,  from “Mar{\'{i}}a de Maeztu” excellence program under project MDM-2016-0692 and CEX2020-001035-M, from Spanish Research State Agency under project PID2020-119632GB-I00. M.D.S. is supported by a startup grant from New Mexico State University. I.V. is supported by the U.S. Department of Energy under Contract No. 89233218CNA000001 and by the LDRD program at LANL.



\bibliography{GLVhydro.bib}

\nolinenumbers

\end{document}